\documentclass[aps,pra,reprint,superscriptaddress,showkeys,nofootinbib,floatfix]{revtex4-1}

\usepackage[T1]{fontenc}
\usepackage{siunitx}
\usepackage{graphicx}
\usepackage{amsmath}
\usepackage{hyperref}
\usepackage{braket}
\usepackage{epstopdf}

\hypersetup{
  colorlinks   = true,
  urlcolor     = blue, 
  linkcolor    = red, 
  citecolor   = blue
}

\begin{document}

\title{Electron cascade for spin readout}

\author{Cornelis~J.~van Diepen}
\author{Tzu-Kan~Hsiao}
\author{Uditendu~Mukhopadhyay}
\affiliation{QuTech and Kavli Institute of Nanoscience,
Delft University of Technology, 2600 GA Delft, The Netherlands}
\author{Christian~Reichl}
\author{Werner~Wegscheider}
\affiliation{Solid State Physics Laboratory, ETH Z\"urich, 8093 Z\"urich, Switzerland}
\author{Lieven~M.~K.~Vandersypen}
\affiliation{QuTech and Kavli Institute of Nanoscience,
Delft University of Technology, 2600 GA Delft, The Netherlands}

\date{\today}

\begin{abstract}
Electrons confined in semiconductor quantum dot arrays have both charge and spin degrees of freedom. The spin provides a well-controllable and long-lived qubit implementation~\cite{Loss1998, Vandersypen2017}. The charge configuration in the dot array is influenced by Coulomb repulsion, and the same interaction enables charge sensors to probe this configuration~\cite{Hanson2007}. Here we show that the Coulomb repulsion allows an initial charge transition to induce subsequent charge transitions, inducing a cascade of electron hops, like toppling dominoes. A cascade can transmit information along a quantum dot array over a distance that extends by far the effect of the direct Coulomb repulsion. We demonstrate that a cascade of electrons can be combined with Pauli spin blockade~\cite{Barthel2009} to read out spins using a remote charge sensor. We achieve $> 99.9\%$ spin readout fidelity in \SI{1.7}{\micro\second}. The cascade-based readout enables operation of a densely-packed two-dimensional quantum dot array with charge sensors placed at the periphery. The high connectivity of such arrays greatly improves the capabilities of quantum dot systems for quantum computation and simulation.
\end{abstract}

\maketitle

Fault-tolerant quantum computation benefits from high connectivity, and requires fast and high-fidelity readout~\cite{Fowler2009}. Qubit connectivity and density are severely limited when charge sensors need to be placed near all quantum dots in the qubit array. Not only the charge sensors themselves take space, but in addition they require a nearby electron reservoir which takes even more space. Several proposals for quantum processors based on gate-defined quantum dots, suggest gate-based readout of two-dimensional arrays to overcome this limitation~\cite{Veldhorst2017, Vandersypen2017, Li2018, Buonacorsi2019}. The comparatively low signal-to-noise ratio (SNR) of this approach has hindered reaching the fidelity required for fault-tolerant quantum computation~\cite{Pakkiam2018, Urdampilleta2019, West2019, Zheng2019}. Signal enhancement has been achieved with a latching scheme~\cite{Studenikin2012, Nakajima2017, Harvey-Collard2018}, but does not enable the readout of dots far from the sensor. 

We show that charge information can be transferred along a quantum dot array with a cascade, in which the spin-dependent movement of one electron induces the subsequent movement of other electrons. Cascades are used in various fields and technologies: stimulated emission~\cite{Einstein1916} in lasers, secondary emission~\cite{Kollath1956} in photomultiplier tubes, impact ionization in avalanche photodiodes~\cite{Webb1974}, and neutron induced decay in nuclear fission~\cite{vonHalban1939}. A cascade has also been used to build classical logic with molecules in scanning-tunneling microscopes~\cite{Heinrich2002} and with excess electrons in cellular automata based on Al islands~\cite{Lent1993, Amlani1999}. 

\begin{figure*}
   \centering
    \includegraphics[scale=.63]{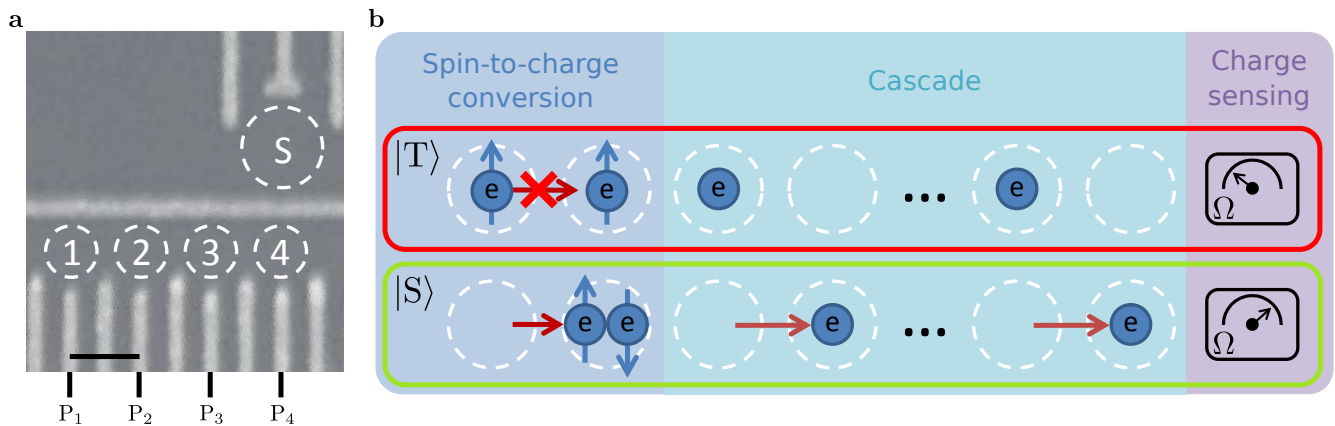}
   \caption{\textbf{Device and cascade concept.} \textbf{a} Scanning electron micrograph of a device nominally identical to the one used for the experiments. Dashed circles labelled with numbers indicate quantum dots in the array, and the dashed circle labelled with "S" is the sensing dot. The scale bar corresponds to \SI{160}{\nano\meter}. \textbf{b} Schematic illustration of spin-to-charge conversion combined with a cascade for electron spin readout on dots far from the charge sensor.}
   \label{fig:device}
\end{figure*}

The prototype for cascade-based readout with quantum dots consists of a quadruple dot and a sensing dot. A scanning electron micrograph image of a device similar to the one used in the experiment is shown in Fig.~\ref{fig:device}a. The device is operated at \SI{45}{\milli\kelvin} and without an external magnetic field, unless specified otherwise. By applying voltages on the electrodes on the surface we shape the potential landscape in a two-dimensional electron gas \SI{90}{\nano\meter} below, formed in a silicon-doped GaAs/AlGaAs heterostructure.
The plunger gates, labelled with $P_i$, control the electrochemical potentials of the dots, and the barrier gates control the tunnel couplings between dots or between a dot and a reservoir. 

Figure~\ref{fig:device}b schematically illustrates the cascade-based readout concept. The first step of the protocol is to perform spin-to-charge conversion, based on Pauli spin blockade (PSB), which induces an initial charge transition conditional on the spin state of the two leftmost electrons. This transition induces a chain reaction of charge transitions with a final charge transition nearby the sensor, which results in a large change in sensor signal. Resetting the cascade can be achieved by undoing the initial charge transition. 

\begin{figure}
   \centering
    \includegraphics[width=.47\textwidth]{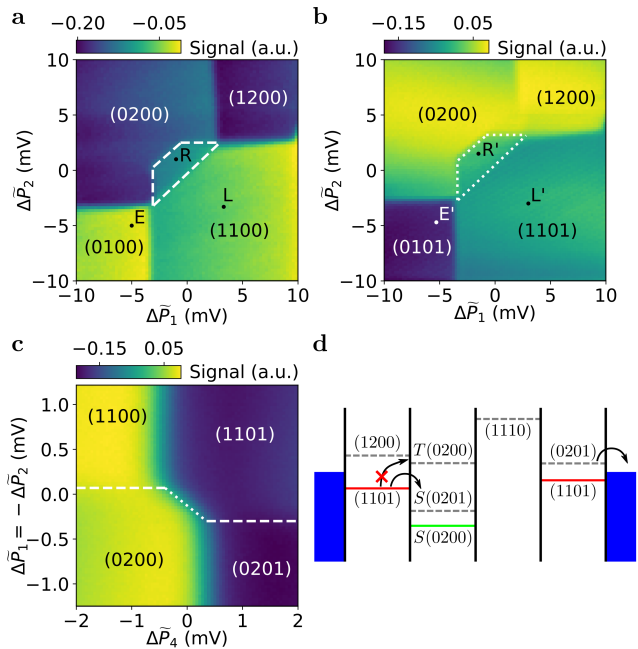}
   \caption{\textbf{Quantum dot tuning for cascade-based spin readout.} Numbers in round brackets indicate charge occupations of the dots. \textbf{a} Charge-stability diagram with transitions for dots 1 and 2. The white, dashed trapezoid on the top-left side of the inter-dot is the Pauli spin blockade (PSB) readout region. The black dots indicate the voltages for the PSB readout cycle: E(mpty), L(oad) and R(ead). \textbf{b} Charge-stability diagram similar to \textbf{a}, but with different occupations of the rightmost dot. The white, dotted trapezoid is the cascade Pauli spin blockade (CPSB) readout region. White and black dots labelled with E$'$, L$'$ and R$'$ indicate the pulse positions for the CPSB readout cycle. Note that the voltages at the origin are different from those in \textbf{a}. \textbf{c} Charge-stability diagram showing both the charge states for PSB and CPSB readout. Dashed (dotted) lines correspond to the charge transitions for the PSB (CPSB) readout regions. \textbf{d} Ladder diagram illustrating the alignment of the dot electrochemical potentials for the CPSB at the readout point. For a triplet state, the system remains in (1101) (red), whereas for as singlet state it transitions to (0201) and then (0200) (green).}
   \label{fig:csd}
\end{figure}

Figure~\ref{fig:csd}a shows a charge-stability diagram with transitions for the two dots on the left. Unless specified differently, the sensing dot is operated on the low-voltage flank of a Coulomb peak throughout this work. For the tuning and measurements, virtual plunger gates $\widetilde{P_i}$ were used for electrochemical potentials~\cite{Hensgens2017, Volk2019, Mills2019} and virtual barrier gates for tunnel couplings~\cite{Hensgens2017, Qiao2020, Hsiao2020}. The charge occupation of the four dots is indicated by the numbers in round brackets. The voltages were swept rapidly from right to left (left to right in panel c) and slowly from bottom to top. With these sweep directions, a white trapezoid caused by PSB is visible to the top-left of the inter-dot transition in the charge-stability diagram, with the sensor signal in between the signal for the $(1100)$ and $(0200)$ charge regions. The trapezoid is the region suited for PSB readout. The distance between the inter-dot line, which is the base of the trapezoid, and the top of the trapezoid, corresponds to the singlet-triplet energy splitting. 

Cascade Pauli spin blockade (CPSB) is seen in the charge-stability diagram of 
Fig.~\ref{fig:csd}b. The fourth dot is tuned close to a charge transition, such that the movement of an electron on the left pair induces a change in charge occupation of the fourth dot. See Supplementary Note~I for details on the tuning of the fourth dot and the sensor for the different charge occupations. Supplementary Note~II contains an analysis of the anti-crossing sizes.

The charge-stability diagram in Fig.~\ref{fig:csd}c shows both the charge states for PSB and for CPSB readout. This diagram is obtained by varying the detuning of the left pair and the potential of the fourth dot. In this diagram there are three different regimes in $\Delta\widetilde{P}_4$. The left and right regions, with charge transitions indicated with a dashed line, can be used for PSB, with dot 4 unoccupied and occupied respectively. The middle region, with a charge transition indicated with a dotted line, can be used for CPSB. 

The tuning requirements of the dot potentials for CPSB readout can be further understood from the ladder diagram in Fig.~\ref{fig:csd}d. Dot 4 needs to be tuned such that $\mu_4(1101) < 0 < \mu_4(0201)$, with the electrochemical potential defined as $\mu_i(\ldots, N_i,\ldots)=E(\ldots, N_i,\ldots)-E(\ldots, N_i-1,\ldots)$, where $E$ is the energy, $N_i$ is the number of electrons on dot $i$, and the Fermi level in the reservoirs is by convention set to zero. This level alignment corresponds to the middle region in Fig.~\ref{fig:csd}c, while the left (right) region corresponds to $\mu_4(1101)$ and $\mu_4(0201)$ both above (below) the Fermi level. Note that if  $\mu_{2,S}(0201)$ is above $\mu_1(1101)$, the cascade in CPSB readout involves a co-tunnel process (see Supplementary~Note~IV). In an alternative implementation, we also perform CPSB readout with a charge transition between $(1110)$ and $(0201)$ (see data in Supplementary~Note~III).

For single-shot PSB readout, voltage pulses are applied as indicated by the black circles in Fig.~\ref{fig:csd}a. The pulse sequence starts in point E, where the charge occupation is $(0100)$. Then the voltages are pulsed to point L, where an electron is loaded from the reservoir onto the leftmost dot reaching the $(1100)$ charge occupation with random spin configuration. Finally, the voltages are pulsed to the readout point, R, where Pauli spin blockade forces a triplet to remain in the $(1100)$ charge occupation while the singlet transitions to the $(0200)$ charge occupation. 

\begin{figure}
   \centering
    \includegraphics[width=.47\textwidth]{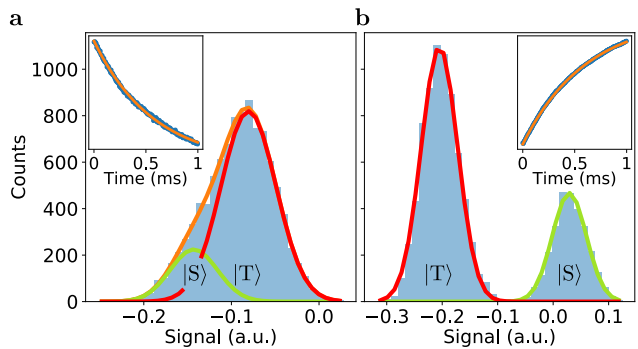}
   \caption{\textbf{Single-shot spin readout.} Histograms and fits of 10,000 single-shot measurements for \textbf{a} PSB readout and \textbf{b} CPSB readout. The integration time is $t_{int}=\SI{1.5}{\micro\second}$. The orange lines are fits to the histograms~\cite{Barthel2009, Zheng2019} and red and green solid lines correspond to respectively the triplet and singlet probability distributions. The left (right) inset shows the signal averaged over the PSB (CPSB) single-shots as a function of wait time in the readout point, and an exponential fit to the data. \textbf{a} For PSB readout the singlet corresponds to charge occupation (0200) and the triplet to (1100). \textbf{b} For cascade PSB readout the singlet also corresponds to (0200) but the triplet corresponds to (1101), thus with an electron on dot 4.}
   \label{fig:signal}
\end{figure}

In Fig.~\ref{fig:signal}a the results of 10,000 single-shot measurements are shown in a histogram. The integration time is $t_{int} = \SI{1.5}{\micro\second}$. The peak at lower sensor signal corresponds to the $(0200)$ charge occupation, and is assigned as singlet, while the peak at higher sensor signal corresponds to the $(1100)$ charge occupation, which is assigned as a triplet. Residual overlap between the singlet and triplet distributions induces errors in the distinction of the two charge states, resulting in errors in the spin readout. The inset shows the signal averaged over the single-shot measurements as a function of the time stamp of the integration window. From an exponential fit, the relaxation time, $T_1 = \SI{724 \pm 70}{\micro\second}$, is obtained (see Supplementary~Note~V).

For CPSB readout, a pulse cycle similar to that for PSB is used. The sensing dot is operated with comparable sensitivity as for PSB readout. The pulse voltages are indicated with white and black circles in Fig.~\ref{fig:csd}b. The pulse sequence again consists of empty, E$'$, load, L$'$, and readout, R$'$. For CPSB, the charge occupation in E$'$ is $(0101)$, and in L$^{\prime}$, again an electron is loaded on the left dot forming the charge state $(1101)$ with a random spin configuration. At the readout point, due to Pauli spin blockade, the two electrons on the left remain on separate dots if they are in a triplet state, which results in the charge state $(1101)$. When the two electrons on the left form a singlet state the resulting charge state will be $(0200)$, because the electron on the left dot moves one dot to the right, and the electron on the fourth dot is pushed off due to the cascade effect (here $\mu_1(1101) > \mu_{2,S}(0201)$, so the two charge transitions can occur sequentially, as discussed above).

Figure~\ref{fig:signal}b shows a histogram of 10,000 CPSB single-shot measurements. The integration time, \SI{1.5}{\micro\second}, is the same as for the PSB single-shot data. The peak at lower sensor signal corresponds to the $(1101)$ charge state, and is assigned as triplet, while the peak at higher sensor signal corresponds to the $(0200)$ charge state, which is assigned as singlet. The residual overlap between the singlet and triplet distributions is strongly reduced for CPSB as compared to PSB. Again, from an exponential fit to the averaged single-shot measurements (inset Fig.~\ref{fig:signal}b), the relaxation time, $T_1 = \SI{680 \pm 3}{\micro\second}$, is obtained. 

The cascade enhances the signal-to-noise ratio for distinguishing between the singlet and triplet states by a factor of $3.5$, extracted by comparing the histogram of CPSB to that of PSB. The SNR is defined as $|V_T - V_S|/\bar{\sigma}_{FWHM}$, with $V_T$ and $V_S$ the signals for a triplet and singlet state respectively, and $\bar{\sigma}_{FWHM}$ the average of the full width at half maximum of the singlet and the triplet probability distributions. Furthermore, Fig.~\ref{fig:signal} shows that for PSB the charge signal for the singlet is lower than that for a triplet, while for CPSB the charge signal for a singlet is actually higher than for a triplet.

The enhanced SNR for CPSB readout arises from two contributions. The first contribution is directly due to the cascade, which maps a charge transition far from the sensor to a charge transition nearby the sensor. The longer the cascade, the larger the relative difference, because the final charge transition remains close to the sensor, while the initial transition is further away for a longer cascade, thus inducing a weaker sensor signal. The second contribution to the SNR enhancement is because the initial charge transition is an inter-dot transition, while the final transition induced by the cascade is a dot-reservoir transition, which has a stronger influence on the sensor. 

As for which spin state produces the highest charge signal, for the case of PSB the singlet signal corresponds to a charge moving closer to the charge sensor, thus the sensor signal goes down. For CPSB a singlet outcome also causes a charge to move closer to the charge sensor, but on top of that a charge is pushed out of the fourth dot, reducing the total charge on the dot array and removing a charge which was very close to the sensor. In this case the two contributions to the signal partially cancel each other, but the resulting effect on the charge sensor is still stronger for CPSB than for conventional PSB. In Supplementary~Note~III CPSB is implemented such that the charge transition induced by the cascade corresponds to an electron moving closer to the sensor, by having an electron move from dot 3 to dot 4. In this case the signal was enhanced by a factor of $3.1$ as compared to PSB. Here the effects on the charge sensor of the initial and the final charge transitions add up, but there is no second contribution to the signal enhancement as there is not a mapping of an inter-dot transition to a dot-reservoir transition.

The average spin readout fidelity for singlet and triplet with CPSB is above $99.9\%$, and is achieved within the \SI{1.7}{\micro\second} readout time. The fidelity for conventional PSB with the same integration time is $85.6 \%$. These fidelities are obtained by analysing the different error sources: relaxation, excitation, and non-adiabaticity. Here, we provide details on the analysis for CPSB (see Supplementary~Note~VI for details on PSB). The residual overlap between the singlet and triplet charge signals and relaxation events during the integration time result in an error of $\eta_{hist} = 0.068\%$ for the average readout fidelity~\cite{Barthel2009, Zheng2019}, as determined from the fit to the single-shot histogram. Relaxation during the arming time, $t_{arm} = \SI{0.2}{\micro\second}$, contributes an error of $0.015\%$. During the arming time, which is the time between the start of the readout pulse and the start of the integration window, the signal is not analysed as it is still rising due to the limited measurement bandwidth. Excitation during the arming and integration time causes an error in the average readout fidelity of $0.014 \%$, with the excitation time, $T_{exc}=\SI{6.0 \pm .3}{\milli\second}$ (see Supplementary Note~V). Another error occurs if the initial charge transition expected for the singlet does not occur, due to charge non-adiabaticity (and slow subsequent charge relaxation). We upper bound this error using the Landau-Zener formula, obtaining $10^{-9} \%$. The effect of the hyperfine field on the mapping to the measurement basis is discussed in Supplementary Note~VII. Relaxation during the voltage ramp is negligible, because there are no relaxation hot-spots on the voltage path for the read-out pulse~\cite{Srinivasa2013} and the ramp time is very short (\SI{10}{\nano\second}). Errors due to anti-crossings with leakage states are not present in this experiment, but are a potential source of errors, when an external magnetic field is applied. Supplementary~Note~IV provides an analysis on scaling of the cascade. 

\begin{figure}
   \centering
   \includegraphics[width=.47\textwidth]{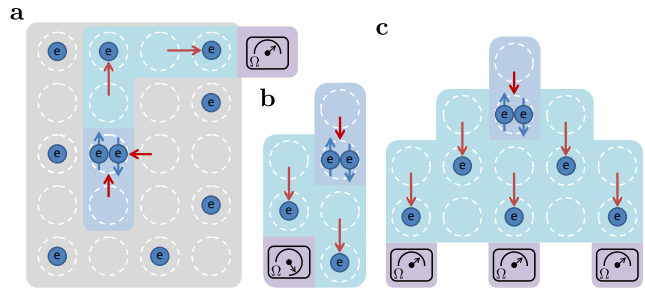}
   \caption{\textbf{Cascade-based readout in 2D and fanout.} Schematic illustrations of cascade-based readout \textbf{a} in a two-dimensional quantum dot array, \textbf{b} using fanout of cascade paths that converge on a single sensor, and \textbf{c} for fanout with multiple sensors. Coloured regions indicate the different aspects of the cascade-based spin readout, with the same colour coding as in Fig.~\ref{fig:device}b. In each panel, the quantum dots are filled in a chequerboard manner prior to readout, but cascades also work with other initial charge configurations. In \textbf{a} an electron is first moved to a dot next to an occupied dot, then PSB is performed.}
   \label{fig:scaling}
\end{figure}

A cascade can also be implemented in large two-dimensional quantum dot arrays. Figure~\ref{fig:scaling}a shows a schematic illustration of an example of cascade-based spin readout in a two-dimensional array. The quantum dots are filled in a chequerboard manner, compatible with the proposal in~\cite{Li2018}, and the sensor is placed at the periphery of the two-dimensional array, with sufficient space for reservoirs. The cascade is implemented by forming a path of dots which are each tuned close to a charge transition, while the dots outside the cascade path are tuned deep in Coulomb blockade so their occupations are unchanged. Cascades can also be designed in a fanout shape, as schematically illustrated in Fig.~\ref{fig:scaling}b. The ends of multiple cascade paths, both triggered by the same initial spin-dependent charge transition, converge at the same sensor, thus increasing the change in charge distribution in the vicinity of the sensor, which increases the SNR and the readout fidelity. Figure~\ref{fig:scaling}c shows another example of fanout of cascade paths, with at the end of each path a sensor. The signal from multiple sensors can be combined to achieve higher SNR and increased readout fidelity. 

We end with a few important considerations on the usefulness of the cascade mechanism for spin readout. First, each electron along a cascade path can itself still be operated as a spin qubit, because phase shifts and reproducible (artificial or natural) spin-orbit induced rotations due to the motion of the electrons can be accounted for in hardware or software~\cite{Watson2018}. Second, as the length of the cascade path increases, both the spin readout fidelity and the timing of the motion of the electrons, can be largely maintained by allowing a cascade to propagate step-by-step using a series of voltage pulses applied to successive dots, see Supplementary Note~IV. Third, the increased SNR from cascades, and the option of further increases through fanout, may enable high-fidelity readout with sensing dots at elevated temperatures~\cite{Yang2019, Petit2019}, because the enhanced signal compensates for the additional thermal noise, allowing higher cooling power and integration with cryogenic control or readout electronics~\cite{Vandersypen2017}. Fourth, the cascade can also be performed with other spin-to-charge conversion methods, for example with energy-selective tunnelling~\cite{Elzerman2004}. Such readout with a cascade does not require a charge sensor nearby the spin to read out, but it does require a nearby reservoir for the initial charge transition.

In conclusion, we have demonstrated a cascade of electrons in a quantum dot array. We combined the cascade with Pauli spin blockade, and achieved spin readout fidelity above $99.9\%$ in \SI{1.7}{\micro\second}, even though the spins were far from the charge sensor. We proposed that a cascade-based readout scheme will enable high-fidelity readout of spins in the interior of a two-dimensional quantum dot array, and that fanout of cascades can be used to enhance the signal further. Other platforms, for example topological qubits, can also benefit from a cascade-based readout, when combined with parity-to-charge conversion~\cite{Aasen2016}. The cascade of electrons opens up a new path for high-fidelity readout in large-scale quantum dot arrays, which is compatible with the established, high-sensitivity, charge sensor, paving the way for further progress in quantum computation and simulation with quantum dot arrays.

\section*{Acknowledgements} 
The authors thank the members of the Vandersypen group for stimulating discussions. This work was supported by the Dutch Research Council (NWO Vici), and the Swiss National Science Foundation.

\section*{Author Contributions}
C.J.v.D. performed the experiment and analysed the data, C.J.v.D. and T.-K.H. conceived the experiment and interpreted the data, U.M. fabricated the device, C.R. and W.W. grew the heterostructure, and C.J.v.D. and L.M.K.V. wrote the manuscript with comments from T.-K.H, and U.M.

\section*{Competing Interests}
The authors C.J.v.D., and T.-K.H. are inventors on a patent application on cascade readout filed by Delft University of Technology (application no. 2024580). The other authors declare that they have no competing interests.

\section*{Additional Information}
Supplementary information for this paper is available at \textbf{link}.

\section*{Methods}
\textbf{Device and set-up} The material for the sample was grown with molecular beam epitaxy and consists of a GaAs/Al$_{0.3}$Ga$_{0.7}$As heterostructure with a silicon doping layer of density \SI{7e12}{\per\centi\meter\squared} at \SI{50}{\nano\meter} depth from the surface. A two-dimensional electron gas (2DEG) was formed at the interface, which is \SI{90}{\nano\meter} below the surface. The mobility was \SI{1.6e6}{\centi\meter^2/\volt\second} at an electron density of \SI{1.9e10}{\per\centi\meter\squared}, measured at \SI{4}{\kelvin}. A single layer of metallic gates (Ti/Au) is defined by electron-beam lithography. The gate pattern was designed to define eight quantum dots and two sensing dots. The device was cooled inside an Oxford Kelvinox 400HA dilution refrigerator to a base temperature of \SI{45}{\milli\kelvin}. To reduce charge noise, the sample was cooled with bias voltages on the gates varying between 100 and \SI{200}{\milli\volt}. Gates $P_1, P_2, P_3$ and $P_4$ were connected to bias-tees ($RC=\SI{470}{\milli\second}$), enabling application of a d.c. voltage as well as high-frequency voltage pulses. Voltage pulses were generated with a Tektronix AWG5014. The sensing dot resistance was probed with radio-frequency reflectometry. The $LC$ circuit for the reflectometry matched a carrier wave of frequency \SI{97.2}{\mega\hertz}. The inductor, $L = \SI{3.9}{\micro \henry}$, was a homebuilt, micro-fabricated NbTiN superconducting spiral inductor, and was wire-bonded to an ohmic contact. The reflected signal was amplified at \SI{4}{\kelvin} with a Weinreb CITLF2 amplifier, and at room-temperature I/Q demodulated to baseband and filtered with a \SI{10}{\mega\hertz} low-pass filter. Data acquisition was performed with a Spectrum M4i digitizer card. After digitization, the I and Q components of the signal were combined with inverse-variance weighting. 

\textbf{High-frequency voltage control}
The voltages for the charge-stability diagrams were simultaneously swept in \SI{78}{\micro\second} for the horizontal direction, and \SI{6.2}{\milli\second} for the vertical direction. The signal was averaged over 1,000 repetitions of such voltage scans. For the single-shot measurements the voltage pulse durations were \SI{100}{\micro\second}, \SI{100}{\micro\second}, and \SI{1}{\milli\second} for respectively the empty, load and read stage. After the read stage a compensation stage of \SI{1}{\milli\second} was performed to prevent accumulation of charge on the bias-tees.

\textbf{Software} The software modules used for data acquisition and processing were the open source python packages QCoDeS, which is available at https://github.com/QCoDeS/Qcodes, and QTT, which is available at https://github.com/QuTech-Delft/qtt.

\textbf{Readout errors}
The error in the average spin readout fidelity, caused by relaxation during the arming time is estimated to be below $\eta_{arm} = \frac{1}{2}\left(1-\exp\left(-t_{arm}/T_1\right)\right) = 0.015\%$. The error due to excitation during the arming and integration time is estimated to be below $\eta_{exc} = \frac{1}{2} (1 - \exp(- (t_{arm}+t_{int}) / T_{exc}) = 0.014 \%$. The error due to charge non-adiabaticity (and slow subsequent charge relaxation) is upper bounded with the Landau-Zener formula to $\eta_{LZ} = \exp\left(-\frac{2 \pi \alpha^2 \Delta t}{\hbar \Delta E}\right) = 10^{-9} \%$, with $\alpha=\sqrt{2} t_{c,12}$, and $t_{c,12} = \SI{11.5}{\micro \electronvolt}$ the tunnel coupling between dots $1$ and $2$, which is obtained from a spin funnel (see Supplementary~Note~VIII), $\Delta t = \SI{10}{\nano\second}$ is the ramp time of the pulse to the readout point and $\Delta E \approx \SI{1}{\milli\electronvolt}$ is the change in double dot detuning from the load to the readout point. 

\section*{Data Availability}
The data reported in this paper, and scripts to generate the figures, are archived at https://doi.org/10.5281/zenodo.3631337.

\end{document}


\begin{centering}
{\Large Supplementary Information for} \\ \vspace{0.2cm}
{\Large \textbf{Electron cascade for spin readout}}\\
\vspace{0.4cm}

{\normalsize Cornelis~J.~van~Diepen$^{1}$, Tzu-Kan~Hsiao$^{1}$, Uditendy~Mukhopadhyay$^{1}$, Christian~Reichl$^{2}$, Werner~Wegscheider$^{2}$, Lieven~M.~K.~Vandersypen$^{1}$}\\
\vspace{0.4cm}
\normalsize{$^{1}$QuTech and Kavli Institute of Nanoscience, Delft University of Technology, 2600 GA Delft, The Netherlands}\\
\normalsize{$^{2}$Solid State Physics Laboratory, ETH Z\"urich, 8093 Z\"urich, Switzerland}\\
\end{centering}

\date{\today}

\tableofcontents

\newpage

\section{Signal in cascade csd}
The signal for Fig.~2b in the main text was taken with the fourth dot and the sensor tuned such that all charge occupations result in clearly distinguishable signals. The signal for $(1101)$ is higher than that for $(0101)$, because the rightmost dot was close to the Fermi level for $(1101)$, and thus only partially occupied. The signal for $(1200)$ is higher than for $(0200)$, because the signal for these occupations is from the high-voltage flank of a sensing dot Coulomb peak. The other relative signals are as intuitively expected, namely adding charges and bringing charges closer to the sensor both result in a reduced sensor signal.

\section{Operating window for cascade readout}
In order to get insight in the size of the operating window for PSB, CPSB and inter-dot CPSB (iCPSB), we start from the single-band Fermi-Hubbard Hamiltonian for the quantum dot array~\cite{Hensgens2017}:
\begin{equation}
H = - \sum_i \epsilon_i n_i + \sum_i \frac{U_i}{2} n_i (n_i - 1) + \sum_{ij, i\neq j} V_{ij} n_i n_j - \sum_{\langle i, j \rangle} t_{c,ij} \left( c^{\dagger}_i c_j + h.c. \right),
\end{equation}
where $\epsilon_i$ is the single-particle energy offset, $n_i = c^{\dagger}_i c_i$ is the dot occupation, and $c^{(\dagger)}_i$ is the annihilation (creation) operator, $U_i$ is the on-site Coulomb repulsion, and $V_{ij}$ the inter-site Coulomb repulsion. For simplicity, we assume in what follows homogeneous Coulomb repulsion, thus $U_i=U$, $V_{i,i+1}=V$, $V_{i,i+2}=V'$, and $V_{i,i+3}=V''$, and neglect tunnel coupling. The shifts of charge transition lines due to capacitive couplings, and the tuning of dot potentials for each of the different readout schemes, namely PSB, CPSB and iCPSB, are obtained by solving sets of constraints.

\subsection{Pauli spin blockade}
The on-site potential $\epsilon_1$ must satisfy $\mu_1(1200) > 0 > \mu_1(1100)$, which yields $2V > \epsilon_1 > V $. The constraint for $\epsilon_2$ follows from $E(1200) > E(1100) > E(0200)$, which yields $U + V > \epsilon_2 > \epsilon_1 + U - V$. From these constraints it follows that the shifts of the relevant charge transition lines due to capacitive couplings are $V$ when projected onto $\epsilon_1$ or $\epsilon_2$. For the two dots on the right to remain empty $ \mu_3(1010) > 0 $ and $ \mu_4(1001) > 0 $, which respectively yield $\epsilon_3 < V'$ and $\epsilon_4 < V''$. 

\subsection{Cascade with dot-reservoir}
The constraint for dot 1 is now $\mu_1(1200) > 0 > \mu_1(1101)$, which yields $2V > \epsilon_1 > V + V''$. We also require $E(1200) > E(1101) > E(0200)$, which yields $ \epsilon_4 + U + V - V' - V'' > \epsilon_2 > \epsilon_1 + \epsilon_4 + U - V - V' - V'' $. For the rightmost dot, the cascade occurs when $\mu_4(0201) > 0 > \mu_4(1101)$, which yields $2V' > \epsilon_4 > V' + V''$. From these constraints it follows that the shifts of the relevant charge transition lines are $V-V''$ when projected onto $\epsilon_1$ or $\epsilon_2$, and $V' - V''$ when projected onto $\epsilon_4$ or onto $\epsilon_1 - \epsilon_2$. For the third dot to remain empty $ \mu_3(1110) > 0$, which yields $\epsilon_3 < V'$.

\subsection{Cascade with inter-dot}
For the leftmost dot $\mu_1(1201) > 0 > \mu_1(1110)$, which yields $2V + V'' > \epsilon_1 > V + V'$. Another requirement is $E(1201) > E(1110) > E(0201)$, which yields $\epsilon_3 - \epsilon_4 + U + V'' > \epsilon_2 > \epsilon_1  + \epsilon_3 - \epsilon_4 + U - 2V + V'$. For the two dots on the right, the cascade effect takes place when $\mu_3(1110) < \mu_4(1101)$, which yields $\epsilon_3 - \epsilon_4 > V - V''$, and $\mu_3(0210) > \mu_4(0201)$, which yields $\epsilon_3 - \epsilon_4 < 2V - 2V'$. In addition, for the rightmost dot, we need $\mu_4(0201) < 0 < \mu_4(1111)$, which yields $2V' < \epsilon_4 < V + V' + V''$. From these constraints it follows that the shifts of the relevant charge transition lines are $V-V'+V''$ when projected onto $\epsilon_1$ or $\epsilon_2$, and $V - 2 V' + V''$ when projected onto $\epsilon_1 - \epsilon_2$ or $\epsilon_3 - \epsilon_4$. 

\section{Inter-dot cascade Pauli spin blockade}
\begin{figure}
   \centering
    \includegraphics[scale=.6]{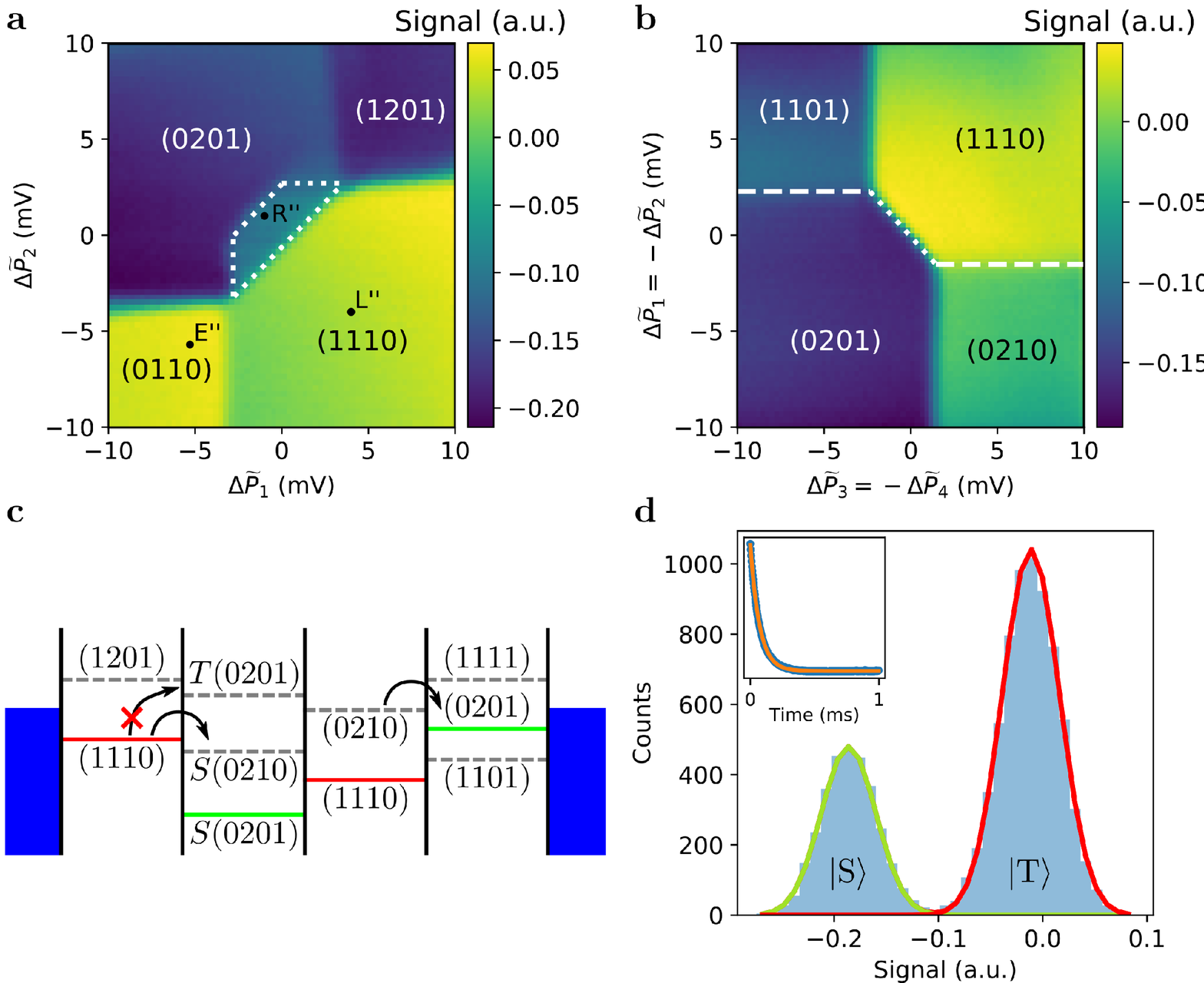}
   \caption{\textbf{Cascade-based readout with inter-dot transition.} Numbers in round brackets indicate charge occupations of the dots. \textbf{a} Charge-stability diagram as a function of the virtual plunger gates of dots 1 and 2. The trapezoid on the top-left side of the inter-dot transition is the inter-dot cascade Pauli spin blockade (iCPSB) window. The black dots indicate the voltages for the iCPSB readout cycle: E$''$(mpty), L$''$(oad) and R$''$(ead). \textbf{b} Charge-stability diagram showing the effect of the inter-dot transition for dots 1 and 2 on the inter-dot transition for dots 3 and 4. On the left and the right, the electron in dot 3 or 4 stays in place when an electron moves from dot 1 to dot 2. In the center, an electron moves from dot 3 to dot 4 when the electron on dot 1 is pushed to dot 2. This corresponds to an inter-dot cascade effect. \textbf{c} Ladder diagram corresponding to the readout point R$''$, illustrating the tuning of the dot potentials for the cascade Pauli spin blockade with inter-dot transition. Note that $\mu_{2,S}(0210)$ is drawn below $\mu_1(1110)$, but for the cascade it could also be above. \textbf{d}  Histograms and fits of 10,000 single-shot measurements for iCPSB readout. The integration time is $t_{int}=\SI{1.5}{\micro\second}$. Red and green solid lines correspond to the respectively triplet and singlet probability distributions, obtained from the fit to the histogram~\cite{Barthel2009,Zheng2019}. For iCPSB readout the singlet corresponds to charge occupation $(0201)$ and the triplet to $(1110)$. The inset shows the signal averaged over the single-shots and an exponential fit, with $T_1 = \SI{75.0 \pm 0.2}{\micro\second}$.}
   \label{fig:icpsb}
\end{figure}

An alternative implementation for cascade-based readout in a quadruple dot is shown in Fig.~\ref{fig:icpsb}. The additional electron moves from the third dot to the fourth dot, thus the cascade involves an inter-dot transition. The signal is from the left flank of a Coulomb peak of the sensing dot. The signal changes of the two charge transitions now add up, thus a singlet state corresponds, as with Pauli spin blockade, to the peak at lower signal and a triplet state corresponds to the peak at higher signal. 

\section{Theory on cascade speed}
In order to assess the scalability of the cascade-based readout, we analyse the speed and adiabaticity of the movement of charges in the cascade. The speed of the cascade is important since spin measurement must be faster than spin relaxation for achieving high-fidelity spin readout. Furthermore, spin readout must be faster than spin decoherence (with dynamical decoupling) for achieving fault-tolerance using feedback in quantum error correction. The adiabaticity with respect to charge is important when the Zeeman splitting is different between quantum dots. For different Zeeman splitting, the uncertainty in the electron position results in a phase error. 

\subsection{Co-tunnel cascade Pauli spin blockade}
When the cascade is operated such that $E(S(0201)), E(1100) > E(1101)$, then the cascade occurs via a co-tunnel process, and the cascade can be operated adiabatically. For a quantum dot array with length four and when the cascade involves a dot-reservoir transition, the relevant charge states are $(1101)$, $(0200)$, $(1100)$ and $(0201)$. The Hamiltonian in this basis is
\begin{equation}
\begin{pmatrix}
- \epsilon_1 - \epsilon_2 - \epsilon_4 + V + V' + V'' & 0 & - t_{c,4R} & - \wt{t}_{c,12}\\
0 & - 2 \epsilon_2 + U & - \wt{t}_{c,12} & - t_{c,4R} \\
- t_{c,4R} & - \wt{t}_{c,12} &  - \epsilon_1 - \epsilon_2 + V & 0 \\
- \wt{t}_{c,12} & - t_{c,4R} & 0 & - 2 \epsilon_2 - \epsilon_4 + U + 2 V'
\end{pmatrix},
\end{equation}
with $\wt{t}_{c,12}=\sqrt{2} t_{c,12}$, and $t_{c,4R}$ the tunnel coupling between the rightmost dot and the right reservoir.
Rewrite the Hamiltonian as
\begin{equation}
\renewcommand*{\arraystretch}{1.5}
\begin{pmatrix}
\frac{\wt{\epsilon}}{2} & 0 & - t_{c,4R} & - \wt{t}_{c,12} \\
0 & - \frac{\wt{\epsilon}}{2} & - \wt{t}_{c,12} & - t_{c,4R} \\
- t_{c,4R} & - \wt{t}_{c,12} & \frac{V'-V''+\wt{\delta}}{2} & 0 \\
- \wt{t}_{c,12} & - t_{c,4R} & 0 & \frac{V'-V''-\wt{\delta}}{2}
\end{pmatrix},
\end{equation}
with $\wt{\epsilon} = \epsilon - U + V + V' + V''$, and $\wt{\delta} = \delta - U + V - 2V'$, where $\epsilon = \epsilon_{12} + \epsilon_4$, and $\delta = \epsilon_{12} - \epsilon_4$, with $\epsilon_{12} = - \epsilon_1 + \epsilon_2$. By diagonalising this Hamiltonian, with approximation $|\wt{\epsilon}| \ll |V'- V'' \pm \wt{\delta}|$, the co-tunnel coupling between the eigenstates that are predominantly (1101) and (0200) is $t_{co} = \frac{\wt{t}_{c,12}t_{c,4R}}{\Delta_+} + \frac{\wt{t}_{c,12}t_{c,4R}}{\Delta_-}$, with $\Delta_{\pm} = \frac{V'-V''\pm\wt{\delta}}{2}$~\cite{Braakman2013b}. 

\subsection{Co-tunnel inter-dot cascade Pauli spin blockade}
The analysis for a cascade involving co-tunnelling and only inter-dot transitions is very similar as for the co-tunnel cascade with a dot-reservoir transition. For cascade with an inter-dot transition, the relevant charge states are $(1110)$, $(0201)$, $(1101)$ and $(0210)$. The Hamiltonian in this basis is
\begin{equation}
\renewcommand*{\arraystretch}{1.5}
\begin{pmatrix}
- \epsilon_1 - \epsilon_2 - \epsilon_3 + 2V + V' & 0 & - t_{c,34} & - \wt{t}_{c,12}\\
0 & - 2 \epsilon_2 - \epsilon_4 + U + 2 V'& - \wt{t}_{c,12} & -t_{c,34} \\
- t_{c,34} & - \wt{t}_{c,12} & -\epsilon_1 - \epsilon_2 - \epsilon_4 + V + V' + V'' & 0 \\
- \wt{t}_{c,12} & - t_{c,34} & 0 & - 2 \epsilon_2 - \epsilon_3 + U + 2V
\end{pmatrix}.
\end{equation}
Rewrite the Hamiltonian as
\begin{equation}
\renewcommand*{\arraystretch}{1.5}
\begin{pmatrix}
\frac{\wt{\epsilon}}{2} & 0 & - t_{c,34} & - \wt{t}_{c,12}\\
0 & - \frac{\wt{\epsilon}}{2} & - \wt{t}_{c,12} & -t_{c,34} \\
- t_{c,34} & - \wt{t}_{c,12} & \frac{V - 2V' + V'' + \wt{\delta}}{2} & 0 \\
- \wt{t}_{c,12} & - t_{c,34} & 0 & \frac{V - 2V' + V'' - \wt{\delta}}{2}
\end{pmatrix}.
\end{equation}
with $\wt{\epsilon} = \epsilon - U + 2V - V'$ and $\wt{\delta} = \delta - U - V + V' + V''$, where $\epsilon = \epsilon_{12} + \epsilon_{34}$, and $\delta = \epsilon_{12} - \epsilon_{34}$ with $\epsilon_{ij} = -\epsilon_i + \epsilon_j$.
By diagonalising this Hamiltonian, with approximation $|\wt{\epsilon}| \ll |V-2V'+V'' \pm \wt{\delta}|$, the co-tunnel coupling between the eigenstates that are predominantly (1110) and (0201) is $t_{co} = \frac{\wt{t}_{c,12}t_{c,34}}{\Delta_+} + \frac{\wt{t}_{c,12}t_{c,34}}{\Delta_-}$, with $\Delta_{\pm} = \frac{V-2V'+V''\pm\wt{\delta}}{2}$.

\subsection{Controlled propagation}
The cascade can be implemented such that the propagation is controlled by a sequence of gate voltages. As example, we consider the cascade Pauli spin blockade as described in the main text. First, conventional PSB is performed with $\mu_4(0201) < 0$. The electron on the fourth dot remains there. Next, gate voltages are changed such that $\mu_4(1101) < 0 < \mu_4(0201)$. Then the cascade will propagate and the electron on the fourth dot will move to the reservoir. A similar scheme can be designed for inter-dot cascade PSB. For a longer cascade path, which involves more than two charge transitions, the propagation could be controlled at each transition. The motivation for controlled propagation becomes clear in the next subsection.

\subsection{Longer cascade}
We now discuss how the total duration of the cascade scales with the length of the cascade path for three different scenarios.

First we consider a cascade where all the charges are displaced in one single co-tunnel process. This involves $N$ simultaneous tunnel events that are each energetically forbidden, but where the final state is lower in energy than the initial state. Then, for a chain with length $2N$ (with every other site occupied, except for the first two sites where the PSB mechanism is implemented), and homogeneous tunnel coupling, $t_{c,ij}=t_c$, and when the cascade involves only inter-dot transitions between neighbouring pairs, the co-tunnel coupling is~\cite{Averin1989}
\begin{equation}
t_{co} = N! \frac{\sqrt{2}t_c^{N}}{2 V^{N-1}},
\end{equation}
where for simplicity we only included inter-site Coulomb repulsion between nearest-neighbour sites. Charge adiabaticity will require increasingly slower gate voltage changes, because $t_c < V$, thus $t_{co}$ decreases exponentially with increasing cascade length. When the adiabaticity condition is not met, the cascade can get stuck along the way.

Next, for the sequential tunneling regime, thus with $E(11LL\ldots L) > E(02LL\ldots L) > E(02RL \ldots L) > \ldots > E(02RR\ldots R)$, with $L=10$ and $R=01$, the expected duration, assuming homogeneous tunnel rates, $\Gamma$, for the individual transitions is~\cite{Buttiker1988}
\begin{equation}
\langle \tau \rangle \sim \frac{N}{\Gamma}.
\end{equation}
For the sequential regime, charge adiabaticity need not be preserved. Charge tunnelling is here a stochastic process and the duration only scales linearly with the length. Note that there is an intermediate regime, which does not fully rely on co-tunnelling, but is also not completely sequential. Theory on this regime is beyond the scope of this work.

Finally, both the charge adiabaticity and speed can be largely maintained in a cascade with controlled propagation. The total duration of the cascade increases linearly with the cascade length, similar to the sequential case, but now uncertainties in the timing of the charge movement can be suppressed, which is important when the Zeeman splittings are not homogeneous along the path.

Alternatively, co-tunnel, sequential, and cascades with controlled propagation could be combined, such that different parts of the cascade have different character. 

\section{Relaxation and excitation time}
The relaxation and excitation time are obtained from the signal averaged over the single-shot measurements, and as a function of time. This signal is fitted with an exponential~\cite{Harvey-Collard2018}
\begin{equation}
V(t) =  A \exp(- \Gamma t) + B,
\end{equation}
with $A$ a pre-factor,
\begin{equation}
\Gamma = \frac{T_1 + T_{exc}}{T_1 T_{exc}},
\end{equation}
and
\begin{equation}
B = \frac{1}{\Gamma} \left( \frac{V_T}{T_{exc}} + \frac{V_S}{T_1} \right),
\end{equation}
where $V_T$ and $V_S$ are obtained from the fit to the histogram of the singlet-shot measurements.

\section{Fidelity analysis for PSB}
From the fit to the histogram in Fig.~3a, the error due to overlap and relaxation during integration is $\eta_{hist} = 14.3 \%$. The relaxation time is $T_1 = \SI{724 \pm 70}{\micro\second}$, which results in an error of $\eta_{arm} = 0.014 \%$. The excitation time is $T_{exc} = \SI{2.8 \pm 1.1}{\milli\second}$, which results in an error of $\eta_{exc} = 0.030 \%$. The error due to charge non-adiabaticity is the same as for CPSB, thus $ 10^{-9} \%$. 

\section{Effect of the hyperfine field}
The measurement basis for spin readout consists of the singlet and triplet states, which are the eigenstates of the Hamiltonian at the readout point. The voltage pulse from the loading point to the readout point, induces a mapping of the eigenstates at the loading point to the measurement basis. This mapping is determined by the pulse duration, the exchange coupling, and the hyperfine field, which is caused by the hyperfine interaction of the electron spins with the nuclear spins. The eigenstates at the loading point can vary between pulse cycles, due to fluctuations of the hyperfine field, thus changing the mapping to the measurement basis. In order to avoid unpredictable mappings, the exchange interaction must dominate the Hamiltonian. This can be done by increasing the exchange interaction or suppressing the hyperfine field fluctuations, either by feedback mechanisms based on dynamical nuclear polarization \cite{Bluhm2010} or by using isotopically purified $^{28}$Si\cite{Veldhorst2014}.

\section{Spin funnel}
\begin{figure}
   \centering
   \includegraphics[scale=.55]{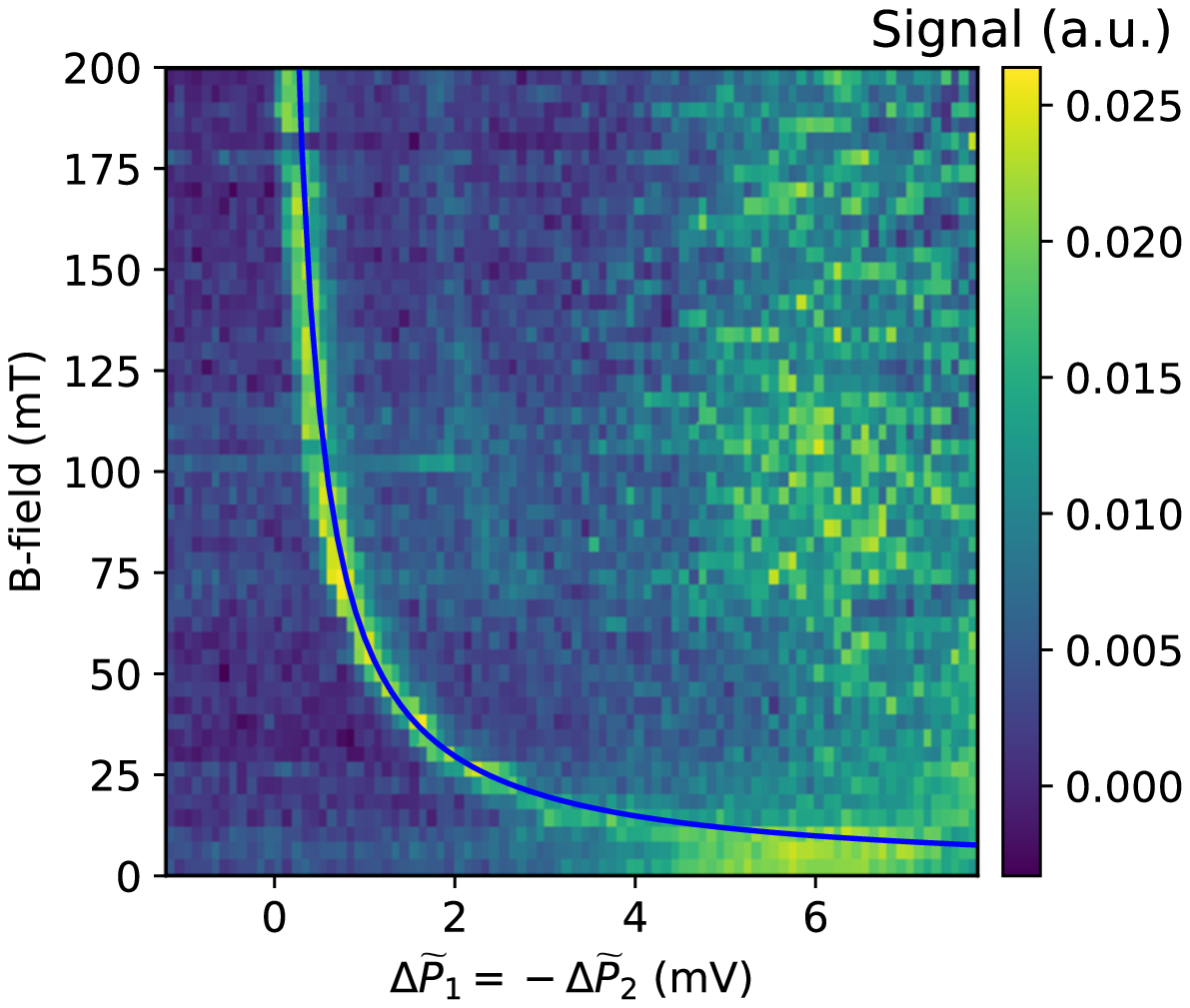}
   \caption{\textbf{Spin funnel} Higher signal corresponds to a higher triplet probability. For each data point the signal is averaged over 1,000 single-shot measurements. The wait time at the operating point, which is in the $(1100)$ charge configuration for $\Delta \wt{P}_1 >0$, was \SI{200}{\nano\second}. The blue, solid line is a fit to $\frac{1}{2}\left(- \epsilon_{12} + \sqrt{8t_{c,12}^2 + \epsilon_{12}^2}\right)$, with $\epsilon_{12} = - \epsilon_1 + \epsilon_2$, the detuning, and $t_{c,12}$ the tunnel coupling between dots $1$ and $2$.}
   \label{fig:spin_funnel}
\end{figure}

The strength of the tunnel coupling between dots $1$ and $2$, $t_{c,12}$, is obtained from a so-called spin funnel measurement, which is shown in Fig.~\ref{fig:spin_funnel}. For the spin funnel, a pulse cycle with three stages is executed~\cite{Petta2005}. The first stage is deep in the $(0200)$ charge region to initialise a singlet state. Then the voltages are pulsed towards the $(1100)$ region, and then into the readout region in $(0200)$. Such a pulse cycle is repeated for varying depths in the $(1100)$ region and varying external magnetic fields. The magnetic field is converted to an energy scale with the $g$-factor, $|g|=0.44$, and the Bohr magneton. From a fit to the funnel, the tunnel coupling $t_{c,12} = \SI{11.5}{\micro\electronvolt}$ is obtained. The detuning is obtained from the change in virtual gate voltages by multiplying with the lever arms, which were obtained with photon-assisted tunnelling measurements~\cite{Hsiao2020}.

